\documentclass[%
 reprint,amsmath,amssymb,aps,superscriptaddress]{revtex4-2}
 
\usepackage[dvipsnames]{xcolor}

\usepackage{graphicx}
\usepackage{dcolumn}
\usepackage{bbm}
\usepackage{ulem}

\begin{document}

\preprint{APS/123-QED}

\title{Cosmic voids and 
filaments from quantum gravity}

\author{J.~Ambj\o rn}
\email{ambjorn@nbi.dk.}
\affiliation{{\small{The Niels Bohr Institute, Copenhagen University, Blegdamsvej 17, DK-2100 Copenhagen Ø, Denmark.}}}

\author{Z. Drogosz}%
 \email{zbigniew.drogosz@doctoral.uj.edu.pl}
 \author{J. Gizbert-Studnicki }%
 \email{jakub.gizbert-studnicki@uj.edu.pl}
 \author{A. Görlich}%
 \email{andrzej.goerlich@uj.edu.pl}
 \author{J. Jurkiewicz}%
 \email{jerzy.jurkiewicz@uj.edu.pl}
 \author{D. Németh}%
 \email{nemeth.daniel.1992@gmail.com}
\affiliation{%
{\small{Institute of Theoretical Physics, Jagiellonian University, \L ojasiewicza 11, Kraków, PL 30-348, Poland.}}}%

\date{\today}

\begin{abstract}
Using computer simulations we study the geometry of a typical quantum universe, i.e., the geometry one might expect before a possible period of inflation. We display it using coordinates defined by means of four classical scalar fields satisfying the Laplace equation with non-trivial boundary conditions. The field configurations reveal cosmic web structures surprisingly similar to the ones  observed in the present-day Universe. Inflation might make these structures relevant for our Universe. 

\end{abstract}

\maketitle

\textit{Introduction} – One major unsolved problem in theoretical physics is how to unite the theory of general relativity and quantum mechanics. 
It is hoped that such a unification will allow us to understand physics at the Planck scale, where the assumed quantum nature of gravity 
most likely plays a dominant role.
Furthermore, the idea of an inflationary period in the history of our Universe has taught us that these quantum fluctuations
at or near the Planck scale can, owing
to an exponential growth of the size of the Universe, freeze and be expanded into seeds for 
macroscopic large-scale structures. Results presented in this article suggest that the scenario of cosmic filaments and voids that we observe in the Universe today might be much more generic in quantum theories of geometry. 

\vspace{4pt}

\noindent
\textit{Lattice Quantum Universes} – In order to discuss  universes of the 
size of a few Planck lengths and their fluctuating quantum geometry one needs a non-perturbative model of quantum gravity. 
We will discuss here a particular model called Causal Dynamical Triangulations (abbreviated to CDT; see \cite{physrep} for a comprehensive introduction and an explanation of its somewhat technical name),
but we believe that our results are generic and will be present in any reasonable quantum model of gravity. 
In CDT, spacetime is a triangulation built by joining together fixed-size four-dimensional simplices in a way that satisfies certain topological requirements.
The edge length $\epsilon$ of the four-simplices acts as an
ultraviolet (UV) cutoff.
Its choice also fixes the geometry of a triangulation. A natural geometric way of calculating the classical Einstein-Hilbert action on such piecewise linear manifolds leads to the so-called Regge action. 
The lattice regularized 
path integral of quantum gravity is then given by
\begin{eqnarray}\label{pathintegral}
\mathcal {Z}_{QG} = \int\mathcal{D}_{\mathcal{M}_H}[g_L]\; e^{iS_{EH}[g_L]} \to \sum_{T_E\in\mathcal{T}_E} e^{- S_R[{T}_E]},
\end{eqnarray}
where  $\mathcal{M}_H$ is a globally defined hyperbolic Lorentzian manifold and $\mathcal{D}_{\mathcal{M}_H} [g_L]$ 
denotes the integration over equivalence classes $[g_L]$ of Lorentzian metrics on $\mathcal{M}_H$.
$\mathcal{T}_E$ is a suitable set of Wick-rotated Euclidean triangulations. The Regge action $S_R[{T}_E]$ for a triangulation $ T_E \in\mathcal{T}_E$ contains the bare couplings related to the cosmological and Newton constants. In principle, we want to adjust the bare coupling constants such that we can take the 
UV cutoff $\epsilon$ 
to zero while keeping physics unchanged (see \cite{renormalization} for a recent review).  
In accordance with the imposed global hyperbolicity, CDT introduces a time foliation of the four-dimensional manifolds into three-dimensional leaves, which are three-dimensional spatial sub-manifolds with a global time $t$ and a fixed topology. 
The explicit CDT construction permits
a Wick rotation of the time coordinate $t$ (still denoted $t$), whereby
CDT becomes a statistical model that can be studied using Monte-Carlo simulations. 
This allows us, \textit{inter alia}, to measure the time dependence of some global quantities, such as the spatial volume. It was shown \cite{semiclassical,c-phase1,c-phase2} that, for suitable choices of the bare coupling constants, both the average spatial volume and its fluctuations can, with a large degree of accuracy, be described by
the Hartle-Hawking minisuperspace model, which assumes isotropy and homogeneity of the Universe such that the only dynamical variable is the scale factor $a(t)$. It should be emphasized that the isotropy and homogeneity is not put in by hand in CDT but follows from integrating out all degrees of freedom other than
the scale factor, and the typical geometries encountered in the quantum path integral 
are not at all close to the classical homogeneous solution of GR. 
The approximate agreement with the classical minisuperspace solution is obtained from an average of an  ensemble of highly fluctuating geometric states and is caused by a non-trivial interplay between the physical action and the entropy of configurations. \\
\indent The CDT model is background independent
and, in the spatial directions, coordinate free.
There is no background geometry in the definition of the path integral. 
However, a good choice of coordinates can be very useful for a description of geometry \cite{coordinates1, coordinates2,long_article}, 
and below we will introduce a coordinate system that is 
suitable 
for describing the highly fluctuating geometries we encounter in the path integral. 


\noindent
\textit{Boundary Conditions} - We will now consider a version of CDT where the piecewise linear manifolds are periodic both in time and space directions. Such a toroidal topology 
can be viewed as a periodically repeated four-dimensional elementary cell, bounded by a set of four independent non-contractible three-dimensional boundaries.
These boundaries are not physical entities, are not unique and
can be locally deformed as long as they still form an elementary cell, and yet they can serve as a reference frame for a coordinate system on a given triangulation.
The non-trivial fractal structure of the encountered geometries makes it difficult to introduce spatial coordinates in a chosen elementary  cell in a constructive geometric way. However, below we will show how to use four massless classical scalar fields, which satisfy the Laplace equation with non-trivial boundary conditions, to parametrize the fractal geometry. 
The fields depend on the geometry, but they do not modify it. One can say that they act as a microscope which uncovers the complicated four-dimensional structure of density fluctuations. Even for a configuration with a very irregular geometry, such fields allow us to define periodic pseudo-continuous coordinates and provide a generalized foliation in all space-time directions. Consequently, it becomes possible to visualize and measure multi-dimensional correlations in all directions.
As will be reported below, what we see is a quantum universe which seems surprisingly similar to our present day macroscopic Universe.

\vspace{4pt}

\noindent
\textit{Scalar fields as coordinates with values on $S^1$ } – We want to find nontrivial harmonic maps between two Riemannian manifolds $\mathcal{M}(g_{\mu\nu}) \to \mathcal{N}(h_{\alpha\beta})$, where $g_{\mu\nu}$ is an arbitrary metric and $h_{\alpha\beta}$ is a flat one. If $\mathcal{N}$ has the topology of $T^4$, then it can be defined by four scalar fields $\phi^\alpha$, $\alpha =1,2,3, 4$, where $\phi^\alpha (x)$ is a map ${\cal M} \to S^1$, such that the following action is minimized: 
\begin{eqnarray}
S_M[\phi,{\cal M}]	= \frac{1}{2} \int \mathrm{d}^{4}x \sqrt{g(x)} \; g^{\mu \nu} (x) \; h_{\rho \sigma}(\phi^\gamma(x)) \nonumber \\
\times \partial_\mu \phi^\rho(x) \partial_\nu \phi^\sigma(x)).
\label{classical_field_eq}
 \end{eqnarray}
Because we have chosen the trivial metric $h_{\rho\sigma}$ on $\mathcal{N}$, eq.\ (\ref{classical_field_eq}) splits up into four independent equations for the four scalar fields $\phi^\sigma$. Minimizing eq.\ (\ref{classical_field_eq}) leads to the set of Laplace equations:
\begin{eqnarray}
\Delta_x \phi^\sigma(x) = 0,\quad 
\Delta_x = \frac{1}{\sqrt{g(x)}}\partial_\mu \sqrt{g(x)}g^{\mu\nu}(x)\partial_\nu.
\label{laplace_eq}
 \end{eqnarray}
Thus $\phi^\sigma$ becomes a harmonic map $\mathcal{M} \to (S^1)^4$. 

Let us consider a trivial one-dimensional example. In this case let $\mathcal{M}$ be $S^1$ with a unit circumference and a positive and strictly periodic density $\sqrt{g(x)}$. 
We want  $x\to \phi(x)$ to be a  non-trivial map $S^1\to S^1$ such that $\phi$ can serve as a coordinate instead of $x$. 
One way to implement this is to find a solution
satisfying 
\begin{equation}
\phi(x+n)=\phi(x)+n \delta,    
\label{phi_cont_eq}
\end{equation}
which maps the circle with a unit circumference to a circle with a circumference $\delta$. The solution to the Laplace equation in this case satisfies
\begin{equation}
 \mathrm{d} \phi(x) = \delta \cdot \sqrt{g(x)} \; \mathrm{d}x.
\end{equation}
By rescaling the field, we can always enforce $\delta=1$. 
The solution $\phi(x)$ is fixed by picking $x_0$ where $\phi(x_0)=0$. The map $x\to \phi(x)$ becomes a monotonically increasing invertible map in the whole domain $\mathbb{R}$. If we parametrize the one-dimensional manifold $\cal M$ in terms of $\phi$ instead of $x$, we will find the volume density in the range $\phi$ to $\phi+\mathrm{d}\phi$ to be proportional to $\sqrt{g(x)}\;\mathrm{d}x$, 
so that effectively $g(\phi) = 1$.
We can also consider a function $\psi(x) =  \textrm{mod}(\phi(x)-\phi(x_1),\delta)$. This function satisfies the Laplace equation in the range between $x_1$ (where $\psi(x)=0$) and $x_1+1$ (where $\psi(x)=\delta$). The equation satisfied by $\psi(x)$ becomes a Poisson equation with the extra inhomogeneous local term, producing jumps at boundary points $x=x_1$ and $x=x_1+1$. It can still be considered to be a Laplace equation with a non-trivial boundary ``jump'' condition.  Generalizing this to  ${\cal M}$ with the topology of $T^4$, we want a solution to the Laplace eq.\ (\ref{laplace_eq}) that wraps around $S^1$ in a particular direction once, and, in addition, we want the points $x$ in ${\cal M}$ that satisfy $\phi^\sigma(x) =c$ to form hypersurfaces $H^\sigma(c)$ whose union for $c$ varying in a range of length 1 covers the whole ${\cal M}$.

\vspace{4pt}

\noindent
\textit{Classical scalar fields with a jump} – In CDT, the four-dimensional manifolds are represented by regular four-dimensional triangulations constructed by gluing together four-simplices so that each face is shared by exactly two simplices.
Denote the number of four-simplices in the triangulation by  $N_4$. Each triangulation we consider is generated by a Monte Carlo simulation, using the CDT partition function. We call such a generated triangulation a configuration. For each configuration we keep information about the position of the four boundaries of the elementary cell. This information
is non-dynamical: it does not influence the Monte Carlo process. Each boundary is a connected set of three-dimensional
faces, each of which
separates two simplices, 
for instance $i$ and $j$, belonging to the two neighboring elementary cells. The connection $i\to j$ can have either positive or negative orientation, depending on the direction in which we cross the boundary. The boundary between the neighboring elementary cells in a direction $\sigma$ can be parametrized by the $N_4\times N_4$ anti-symmetric matrix $\mathbf{B}_{ij}^\sigma=-\mathbf{B}_{ji}^\sigma$ with the elements
\begin{equation}
\label{adjacency}
\mathbf{B}_{ij}^\sigma =
\begin{cases}
	\pm 1 & \textrm{if $i\to j$ crosses the boundary,}\\
	0 & \textrm{otherwise}.
\end{cases}
\end{equation}
The number of directed boundary faces of a simplex $i$ is given by $b_i^\sigma=\sum_j \mathbf{B}_{ij}^\sigma$,
with the obvious constraints
$-5 < b_i^\sigma < 5$ and $\sum_i b_i^\sigma = 0$. For any simplex $i$ adjacent to a boundary, the values $\mathbf{B}_{ij}^\sigma$ are all positive or zero (on one side of the boundary), or all negative or zero (on the other side). We consider four scalar fields $\phi_i^\sigma$ located in the centers of simplices and solve the minimization problem for the following discrete version of the continuous action in eq.\ (\ref{classical_field_eq}), for each field $\phi_i^\sigma$:
\begin{eqnarray}
\label{CDT_laplace_eq}
S_M^{CDT}[\phi^\sigma,{T_E}]= \frac{1}{2} 
\sum_{i \leftrightarrow j} (\phi_i^\sigma - \phi_j^\sigma -\delta \mathbf{B}_{ij}^\sigma)^2 .
\end{eqnarray}
In (\ref{CDT_laplace_eq})  the sum is over all pairs of neighboring four-simplices in the triangulation $T_E$ representing  the manifold $\mathcal{M}(g_{\mu\nu})$ in eq.\ (\ref{classical_field_eq}).
The parameter $\delta$ plays the same role as in the one-dimensional example considered previously,
and here too, by
rescaling the field, we can always set $\delta=1$.
The action (\ref{CDT_laplace_eq}) has two important symmetries.
The first one is the invariance under a constant shift of the scalar field  (the Laplacian zero mode). The second is a local invariance under a modification of the boundary $\mathbf{B}_{ij}^\sigma$ and a shift by $\pm 1$ (depending on the side of the boundary) of the field value in a simplex $i$ adjacent to the boundary. This is equivalent to moving the simplex to the other side of the boundary and compensating for the change of the field in its center.
After such a move, the number of faces belonging to the boundary will in general be changed, but the action (\ref{CDT_laplace_eq}) will remain constant.

The classical field, henceforth 
denoted as $\phi_i^\sigma$,
minimizes the action (\ref{CDT_laplace_eq}),
and thus has to satisfy the non-homogeneous Poisson-like equation
\begin{equation}
\label{classical}
    \mathbf{L}\phi^\sigma =  b^\sigma,
\end{equation} 
where $\mathbf{L}=5 \mathbbm{1} - \mathbf{A}$ is the $N_4\times N_4$ Laplacian matrix, and $\mathbf{A}_{ij}$ is the adjacency matrix with entries of value 1 if simplices $i$ and $j$ are neighbors and 0 otherwise.

The Laplacian matrix $\mathbf{L}$ has a constant zero mode but can be inverted if we fix a value of the field $\phi_{i_0}^\sigma=0$ for an arbitrary simplex $i_0$. Although $\mathbf{L}$ is a sparse matrix, inverting it is  a major numerical challenge for a system typically of size $N_4 \approx 10^6$. 
After we nevertheless obtain the classical solution $\phi^\sigma$, the multi-dimensional analogue of the one-dimensional function $\psi(x)$ is given by
$\psi_i^\sigma$:
\begin{equation}
    \psi_i^\sigma  = \textrm{mod}(\phi_i^\sigma, 1).
\end{equation}
A new boundary is defined by $\bar{b}^\sigma = \mathbf{L}\psi^\sigma$.
This allows us to reconstruct 
a new three-dimensional hypersurface $H$, separating the elementary cell from its copies in the direction $\sigma$ and characterized by the fact that the field jumps from 0 to 1 when crossing $H$. This hypersurface can be moved to another position if we consider a family of hypersurfaces $H(\alpha^\sigma)$ obtained from
\begin{equation}
\label{alpha}
    \psi_i^\sigma(\alpha^\sigma) = \textrm{mod}(\phi_i^\sigma-\alpha^\sigma, 1),\quad
      \bar{b}^\sigma(\alpha^\sigma) = \mathbf{L}\psi^\sigma(\alpha^\sigma).
\end{equation}
\begin{figure}[t]
\includegraphics[width=0.45\textwidth]{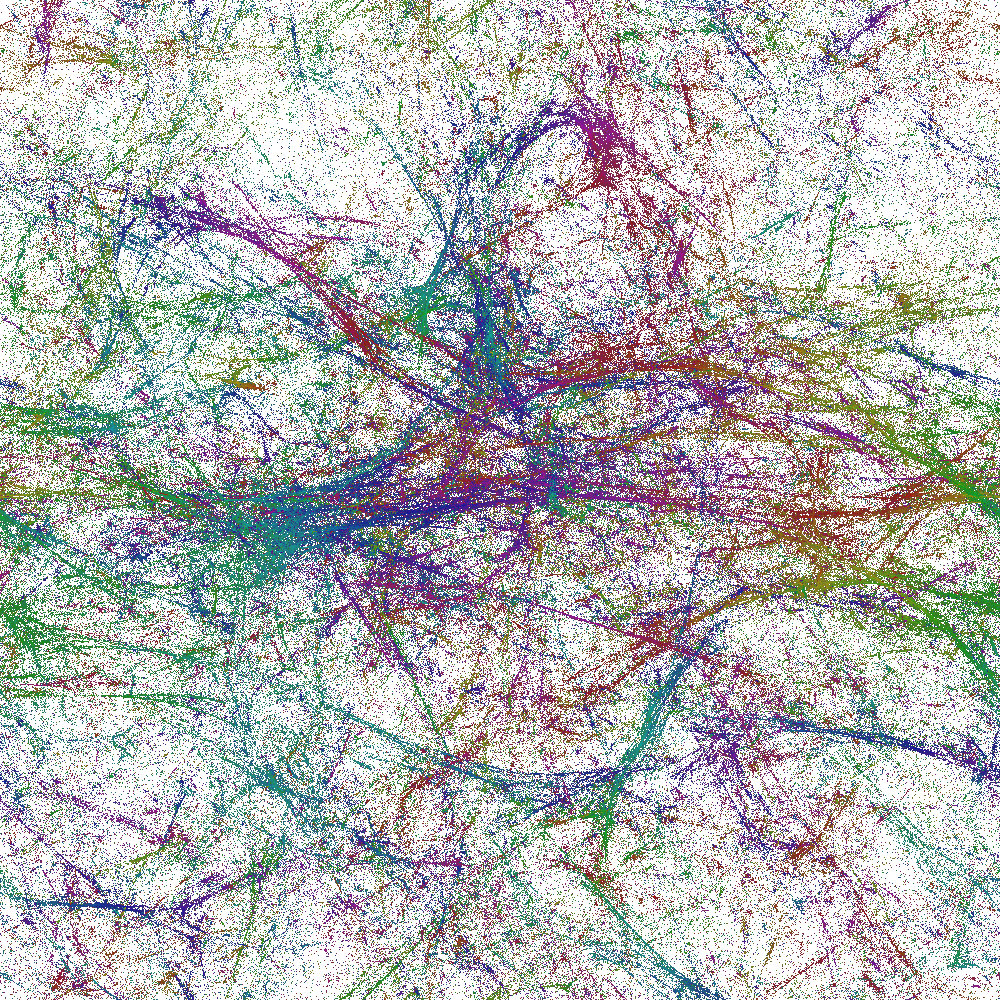} 
\caption{The projection of four-volume on the $xy$-plane, as defined by (\ref{hypers_vol_eq_psi}) for a CDT configuration.
Different colors correspond to different times $t$ of the original $t$-foliation. 
}\label{phi-xy_plot}
\end{figure}
Changing $0\leq \alpha^\sigma<1$, we shift the position of the hypersurface and cover the whole elementary cell defined by the boundary (\ref{alpha}), and in this way we obtain a foliation in the direction $\sigma$. We may now use $\psi_i^\sigma= \psi_i^\sigma(0)$ as a coordinate in the $\sigma$-direction. The same construction can be repeated in all directions $\sigma\in \{ x,y,z,t\}$ for any 
configuration
obtained in the numerical simulations, and in this way every simplex $i$ will be assigned a unique set of coordinates $\{\psi^x_i,\psi^y_i,\psi^z_i,\psi^t_i\}$, all in the range between 0 and 1.
A solution to the Laplace equation 
has the property that coordinates of each simplex are equal to the  mean value of coordinates of its neighbors (up to the shift of the field at the boundary), i.e., it preserves the triangulation structure. This is the required map 
from
our configuration with a topology of $T^4$ to $(S^1)^4$ (which of course also has the topology $T^4$).
Note
that the coordinate $\psi^t_i$ is not the same as the one coming from the original foliation of the CDT model. The parametrization defined above permits to analyze the distribution of the four-volume (the number of simplices)
contained in hypercubic blocks with sizes $\{\Delta\psi^x_i,\Delta\psi^y_i,\Delta\psi^z_i,\Delta\psi^t_i\}$, 
which is equivalent to measuring the integrated $\sqrt{g(\psi)}$:
\begin{eqnarray}
\Delta N(\psi) = \sqrt{g(\psi)} \prod_\sigma \Delta \psi^\sigma = N(\psi) \prod_\sigma \Delta \psi^\sigma.
\label{hypers_vol_eq_psi}
\end{eqnarray}
\begin{figure}[t]
\includegraphics[width=0.45\textwidth]{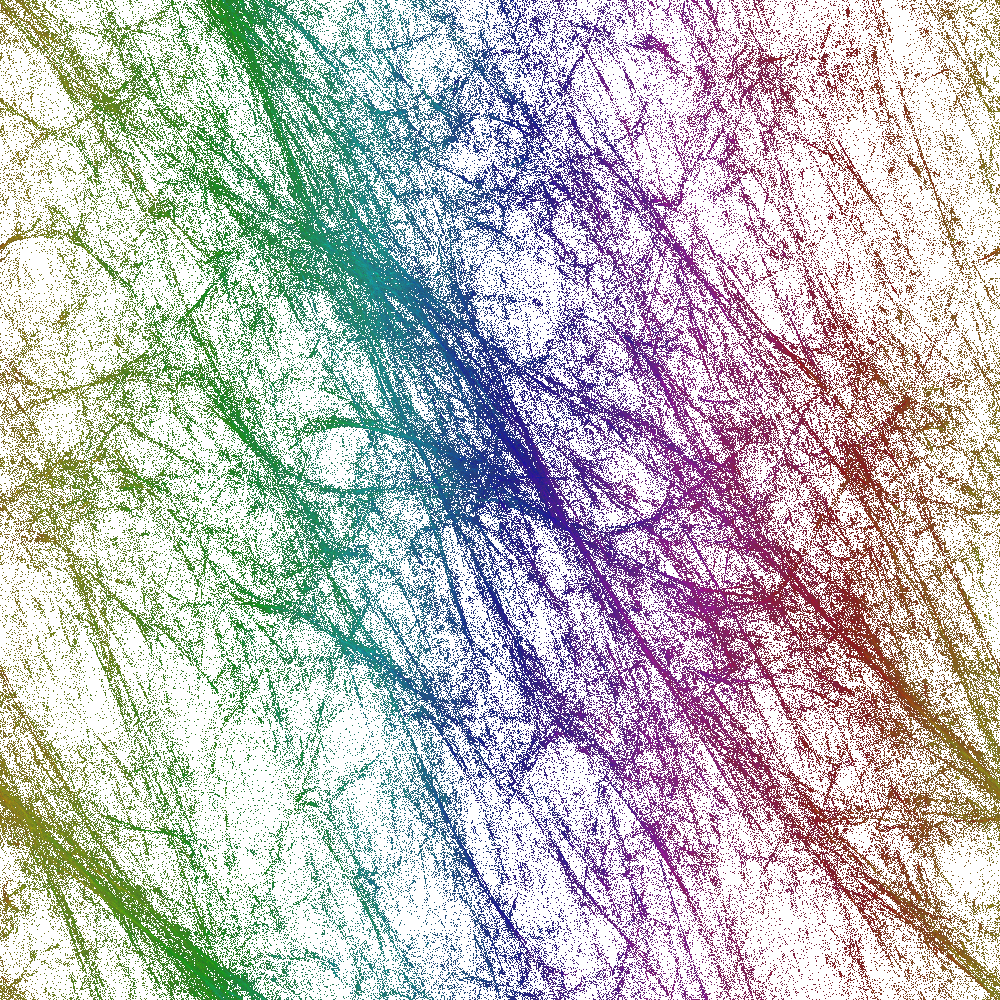}
\caption{
The projection of four-volume on the $tx$-plane  for a CDT configuration.
There is a strong correlation between the original $t$-foliation (color) and new time coordinate $\psi^t$ (horizontal axis).} 
\label{phi-xt_plot}
\end{figure}
We can measure the full four-dimensional distribution $N(\psi)$.
In Figs.\ \ref{phi-xy_plot} and \ref{phi-xt_plot}
we show projections of the volume density distribution of a typical configuration
on two-dimensional parameter subspaces,
the $xy$-plane and the $tx$-plane respectively,
integrating over 
the remaining two directions.
One observes a remarkable pattern of voids and filaments, 
which qualitatively looks quite similar to pictures of voids and filaments observed in our real Universe (see e.g.\ \cite{compare}; the plots can be found on the web \cite{galaxies,figurefrom}). 
Using the new coordinates, we observe a pattern of volume concentrations in the spatial directions. The higher-density domains tend to attract each other, forming denser clouds, which survive in time evolution
(see Fig. \ref{phi-xt_plot}). There seems to appear a sequence of scales, characterizing a gradual condensation of gravitationally interacting ``objects'', but  one should remember that there is no matter in this system, only pure geometry, which behaves as if quantum fluctuations could produce massive interacting objects, somewhat analogous to dark matter. 
Of course, we are talking about quantum objects of Planckian size, but if a more extended model exhibited inflation, one could imagine that aspects of these objects
would be frozen when entering the horizon, like the standard Gaussian fluctuations in simple inflationary models, and then would re-enter the horizon at a later stage, after reheating, as classical densities.

\vspace{4pt}

\noindent
\textit{Discussion} – 
CDT presents us with a model of what we believe are generic fluctuations of geometry at the Planck
scale. Hopefully, 
a detailed analysis of these fluctuations can be used in inflationary models, and such a study is now possible thanks to the coordinates
we introduced here in spacetimes with toroidal topology.
Apart from then being able to discuss the nature of geometric fluctuations before inflation we hope that  measurements of such correlations will allow us to determine ``experimentally'' (i.e., using Monte Carlo data) the effective continuum action that governs our lattice model.
The construction of such an effective action will help us to understand if CDT is an UV-complete quantum field theory of gravity, as imagined in the so-called asymptotic safety scenario, or only an effective quantum theory of geometries.

\vspace{4pt}

\noindent
\textit*{Acknowledgments} – Z.D. acknowledges support from the National Science Centre, Poland, grant 2019/32/T/ST2/00390. J.G.-S. acknowledges support of the grant UMO-2016/23/ST2/00289 from the National Science Centre Poland. A.G. acknowledges support by the National Science Centre, Poland, under grant no. 2015/17/D/ST2/03479. J.J. acknowledges support from the National Science Centre, Poland, grant  2019/33/B/ST2/00589. D.N. acknowledges support from National Science Centre, Poland with grant no. 2019/32/T/ST2/00389.


\begin{thebibliography}{00}
\bibitem{physrep}
 J.~Ambjorn, A.~Goerlich, J.~Jurkiewicz and R.~Loll,
 Phys.\ Rept.\ {\bf 519} (2012) 127\\
 R.~Loll,
 Class.\ Quant.\ Grav.\ {\bf 37} (2020) no.1, 013002
\bibitem{renormalization}
J.~Ambjorn, J.~Gizbert-Studnicki, A.~G\"orlich, J.~Jurkiewicz and R.~Loll,
Front. in Phys. \textbf{8} (2020), 247.
\bibitem{semiclassical}
J.~Ambjorn, J.~Jurkiewicz and R.~Loll,
Phys. Lett. B \textbf{607} (2005), 205-213\\
J.~Ambjorn, A.~Gorlich, J.~Jurkiewicz, R.~Loll, J.~Gizbert-Studnicki and T.~Trzesniewski,
Nucl. Phys. B \textbf{849} (2011), 144-165

\bibitem{c-phase1}
J.~Ambjorn, J.~Jurkiewicz and R.~Loll,
 Phys.\ Rev.\ D {\bf 72} (2005) 064014;
Phys.\ Rev.\ Lett.\ {\bf 93} (2004) 131301.\\
 J.~Ambjorn, A.~G\"orlich, J.~Jurkiewicz and R.~Loll,
 Phys.\ Rev.\ D {\bf 78} (2008) 063544;
Phys.\ Rev.\ Lett.\ {\bf 100} (2008) 091304 

\bibitem{c-phase2}
J.~Ambjorn, Z. Drogosz, J.~Gizbert-Studnicki, A.~G\"orlich, J.~Jurkiewicz and D.~Németh,
Phys. Rev. D {\bf 94} (2016) 044010 \\
J.~Ambjorn, J.~Gizbert-Studnicki, A.~G\"orlich, K.~Grosvenor and J.~Jurkiewicz,
Nucl. Phys. B \textbf{922} (2017), 226-246
\bibitem{coordinates1}
J.~Ambj\o{}rn, Z.~Drogosz, J.~Gizbert-Studnicki, A.~G\"orlich and J.~Jurkiewicz,
Nucl. Phys. \textbf{B} (2019), 114626
\bibitem{coordinates2}
J.~Ambjorn, Z.~Drogosz, A.~G\"orlich and J.~Jurkiewicz,
[arXiv:2007.13311 [hep-th]].
\bibitem{long_article}
J.~Ambjorn, Z. Drogosz, J.~Gizbert-Studnicki, A.~G\"orlich, J.~Jurkiewicz and D.~Németh,
to be published,
\bibitem{compare}
J.N.~Burchett, O.~Elek, N.~Tejos, J.X.~Prochaska, T.M.~Tripp, R.~Bordoloi and
A.G.~Forbes, The Astrophysical Journal Letters, 891:L35 (2020).
\bibitem{galaxies}
https://aasnova.org/2018/01/05/galaxies-growing-up-on-the-edge-of-the-void/

\bibitem{figurefrom}
https://astronomynow.com/2016/08/12/astronomers-use-cosmic-voids-to-study-the-universe/


\end{thebibliography}

\end{document}